\documentclass[3p,times,procedia]{elsarticle}
\usepackage{nupha_ecrc}

\volume{00}

\firstpage{1}

\journalname{Nuclear Physics A}

\runauth{}

\jid{nupha}

\jnltitlelogo{Nuclear Physics A}

\usepackage{amssymb}
\usepackage[figuresright]{rotating}

\begin{document}

\begin{frontmatter}



%% Instructions from Editor: Please use the following \dochead only in the preprint version (e-print arXiv etc.); 
%% use empty \dochead{} when submitting to Nuclear Physics A!
\dochead{XXVIIth International Conference on Ultrarelativistic Nucleus-Nucleus Collisions\\ (Quark Matter 2018)}
%% e.g. \dochead{17th International Conference on Dynamical Processes in Excited States of Solids}

%\title{Hadronic observables in small collisions systems from classical Yang-Mills dynamics + Lund string fragmentation at RHIC}

\title{Hadronic observables in p+p and d+Au collisions at RHIC using CGC+PYTHIA}
%

\author[a]{Bj\"orn Schenke}
\author[b]{S\"oren Schlichting}
\author[a]{Prithwish Tribedy}
\author[a]{Raju Venugopalan}
\address[a]{Physics Department, Brookhaven National Laboratory, Upton, NY 11973, USA}
\address[b]{Department of Physics, University of Washington, Seattle, WA 98195-1560, USA}
%\address[c]{Physics Department, Brookhaven National Laboratory, Upton, NY 11973, USA}
%\address[d]{Physics Department, Brookhaven National Laboratory, Upton, NY 11973, USA}

%\author{}

\address{}

\begin{abstract}
The IP-Glasma model of CGC combined with the Lund model of PYTHIA provides a very successful description of hadron production from gluon dominated non-equilibrium matter in various small collision systems. This new CGC+PYTHIA framework, naturally reproduces several characteristic features of the hadronic observables such as the mass ordering of $v_{2}(p_{T})$ and $\langle p_{T}\rangle$, often ascribed to collectivity driven by hydrodynamics at the LHC~\cite{Schenke:2016lrs}. In this contribution we extend our work to provide a systematic comparison of particle spectra and multiplicity distributions in p+p and d+Au collisions at the RHIC. %We study bulk observables such as particle spectra, nuclear modification factors (R_{pA}), proton-to-pion ratios and multi-particle azimuthal angular correlations. We demonstrate that characteristic features of hadronic observables such as the baryon to meson ratio, mass ordering of v2(pT) and <pT>, are naturally reproduced within the initial state framework [1]. We also present first results on a systematic comparison of such observables across different systems, including p+p and p+Pb collisions at the LHC as well as p/d/He3+Au at RHIC.
\end{abstract}


\end{frontmatter}

% \linenumbers

\section{Introduction}
\label{intro}
The measurements of bulk hadronic observables in small collision systems at the RHIC and the LHC are of prime importance because they provide direct access to the underlying dynamics of multi-particle production in QCD. However, the theoretical description of such observables is challenging because of the dominance of soft non-perturbative processes. The two major challenges are the systematic treatment of the soft multi-parton production and a consistent treatment of the hadronization of soft partons. Although the former is well addressed by the Color Glass Condensate (CGC) effective theory over a wide range of kinematics, an {\it ab inito} QCD based treatment of the latter is still missing. 
%
%This is why CGC computations are 
%
Effective approaches have been developed over the years towards a consistent treatment of hadronization~\cite{Field:1977fa,Azimov:1984np,Andersson:1983ia}. In the phenomenology of A+A collisions, the problem of hadronization is addressed by fluid-dynamic evolution followed by a Cooper-Frye prescription~\cite{Cooper:1974mv}. Applying such a macroscopic approach across systems of all sizes is however challenging~\cite{SchenkeQM18}. The conventional microscopic description of small collision systems such as p+p, on the other hand, implements an effective hadronization model based on the Lund-string fragmentation~\cite{Andersson:1983ia}. Such descriptions are implemented in Monte-Carlo event generators like PYTHIA~\cite{Sjostrand:2014zea}. However, the underlying framework for multi-particle production in PYTHIA fails to describe several observations in high multiplicity events~\cite{SjostrandQM18}. 
Computations based on the CGC approach, on the other hand, explain the origin of such events as a basic feature of rare highly occupied gluon states, as well as provide both qualitative and quantitative description of the most recent observations in such events~\cite{Mace:2018yvl,Schenke:2015aqa}. The only shortcoming over the years in this approach has been that the data-model comparisons are either done at the level of gluon distributions or by employing fragmentation functions~\footnote{Very recently CGC+NRQCD formalism has been used for the hadronization of $J/\psi$~\cite{Ma:2018bax} in high multiplicity p+p/A collisions.}, though the latter are applicable at higher $p_{T}$. 

% ~\footnote{NRQCD very recently using NRQCD formalism.}
%standard parton-hadron independent hadronization  and very recently using NRQCD formalism~\cite{Ma:2018bax}, that are most effective for the hard probes.
%

In \cite{Schenke:2016lrs} an approach was developed by combining the IP-Glasma model of CGC and the Monte-Carlo Lund-string fragmentation of PYTHIA. This new CGC+PYTHIA framework successfully describes bulk observations in p+p collisions like multiplicity distributions $P(n)$ and mean transverse momentum $\left<p_{T}\right>$ of the identified particles. Most importantly, this model naturally describes the systematics of rare high multiplicity events that has recently generated a lot of interest. Observations such as the growth of $\left<p_{T}\right>$ with charged particle multiplicity $N_{ch}$, its mass ordering $\left<p_{T}\right>_{p/\bar{p}}>\left<p_{T}\right>_{K^\pm}>\left<p_{T}\right>_{\pi^\pm}$, the appearance of long-range di-hadron correlations and mass ordering of the elliptic anisotropy coefficient $v_{2}\{2\}$ are often attributed to signatures of collectivity driven by hydrodynamics. The CGC+PYTHIA framework successfully describes such observations without requiring strong final state interactions. 
In this contribution, we employ the CGC+PYTHIA framework and focus on a few basic observables such as multiplicity distributions and transverse momentum spectra in 200 GeV p+p and d+Au collisions.

\begin{figure}
\includegraphics[width=0.45\textwidth]{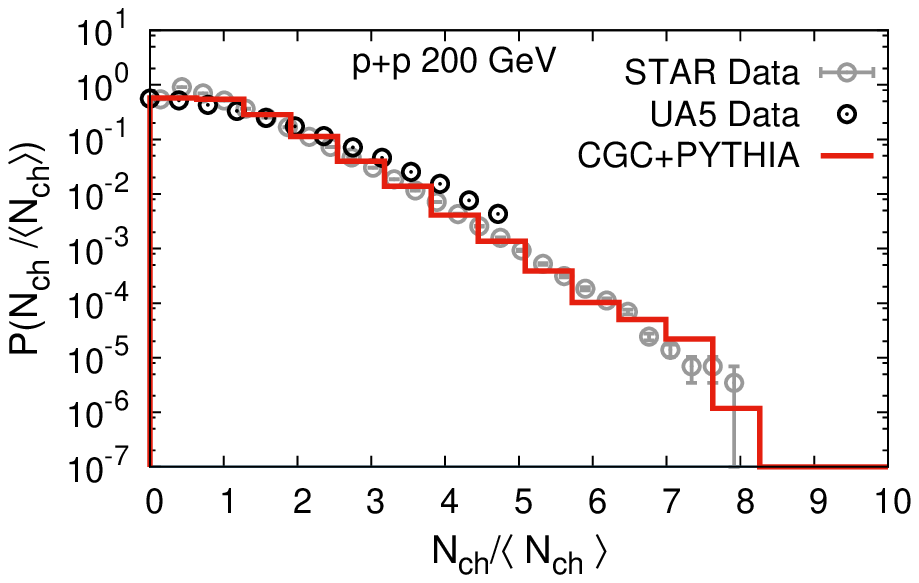}
\includegraphics[width=0.45\textwidth]{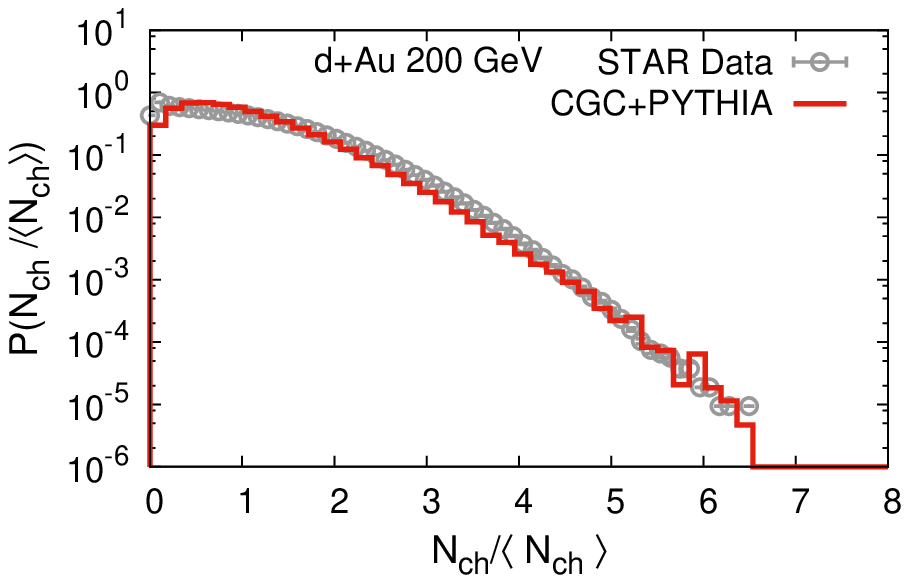}
\caption{(color online) Probability distribution of charged hadron multiplicity compared to STAR~\cite{Abelev:2008ez} and UA5 data~\cite{Ansorge:1988kn}.}
\label{fig_multdist}
%\vspace{-10pt} 
\end{figure}

\section{The CGC+PYTHIA framework}
The details of the CGC+PYTHIA framework are described in~\cite{Schenke:2016lrs}. In this framework, we estimate the event-by-event distribution of gluons $dN_g/dk_\perp dy$ from numerical solutions of the classical Yang-Mills equations as implemented in the IP-Glasma model~\cite{Schenke:2012wb}. The IP-Glasma lattice parameters in our computations are as follows: we use transverse lattices of size N = 400 and spacing a = 0.04 fm; we employ an infrared regulator of mass $m$ = 0.2 GeV.  We choose the ratio of the saturation scale to the parameter controlling the width of the color charge fluctuations to be $Q_s/g^2\mu=0.7$ and the width of the fluctuations of saturation scale $\sigma(\log(Q_S))=0.5$. Other details of the lattice setup are similar to the most recent IP-Glasma computation performed in~\cite{Mantysaari:2018zdd}.  
%
%
%The distribution is boost invariant in rapidity. 
%
%In this work, we perform our computations for p+p and p+Au collisions at RHIC where 
%
We integrate the distribution $dN_g/dk_\perp dy$ to estimate the total number of gluons $N_g$ within the transverse momentum $0\!<\!k_\perp\!<\!k_\perp^{\rm max}$ and rapidity $-y^{\rm beam}\!<\!y\!<\!y^{\rm beam}$, where the beam rapidity at $\sqrt{s}=200$ GeV is $y^{\rm beam} = $5.36.  
We then sample $N_g$ gluons to construct PYTHIA strings and feed them into the Monte-Carlo implementation of the Lund string fragmentation routine in PYTHIA (version 8.235)~\cite{Sjostrand:2014zea}. The Lund symmetric fragmentation function implemented in PYTHIA is given by 
\begin{equation}
f(z, m_{T}) = \frac{1}{z} (1-z)^{a} \exp \left(-\frac{b\, m{_T}^2}{z} \right)\;,
\end{equation}
where the parameters $a,b,m_{T}$ are constrained by global data -- further tuning of the parameters $a$ and $b$ are allowed within the ranges of $0\le a \le2.0$ and $0.2\le b\le2.0$, which we will do below to estimate part of our systematic uncertainties. %In this work we perform variations of $a$ and $b$ within the allowed range to study systematics of our results. 
\section{Results}
In Fig.\ref{fig_multdist} we present the probability distribution of the scaled \footnote{It is worth noting that many theoretical uncertainties, that we discuss later, are mitigated by taking the ratio $N_{\rm ch}/\left<N_{\rm ch}\right>$.} %This also facilitates an approximate data-model comparison when the experimental distribution is uncorrected.}
 inclusive charged hadron multiplicity $P(N_{\rm ch}/\left<N_{\rm ch}\right>)$ at midrapidity in p+p and d+Au collisions compared to STAR~\cite{Abelev:2008ez} and UA5 data~\cite{Ansorge:1988kn}. 
%
 %; the additional advantage is that  experimental STAR data that are uncorrected 
%
%We point out that the STAR data are not corrected for efficiency, however taking the ratio $N_{\rm ch}/\left<N_{\rm ch}\right>$
%There are several sources of sus
We find a nice agreement over the entire range of available data, slightly better than a previous CGC computation that does not include fragmentation~\cite{Mace:2018vwq}. %
It is widely known that $P(N_{\rm ch}/\left<N_{\rm ch}\right>)$ is intrinsically composed of multiple negative binomial distributions. To the best of our knowledge, CGC is the only available framework at the RHIC and LHC that produces such negative binomial distributions from first principles -- conventional models like PYTHIA generate convolution of Poisson distributions for charged hadron multiplicity~\cite{SjostrandQM18}.

\begin{figure}
\includegraphics[width=0.45\textwidth]{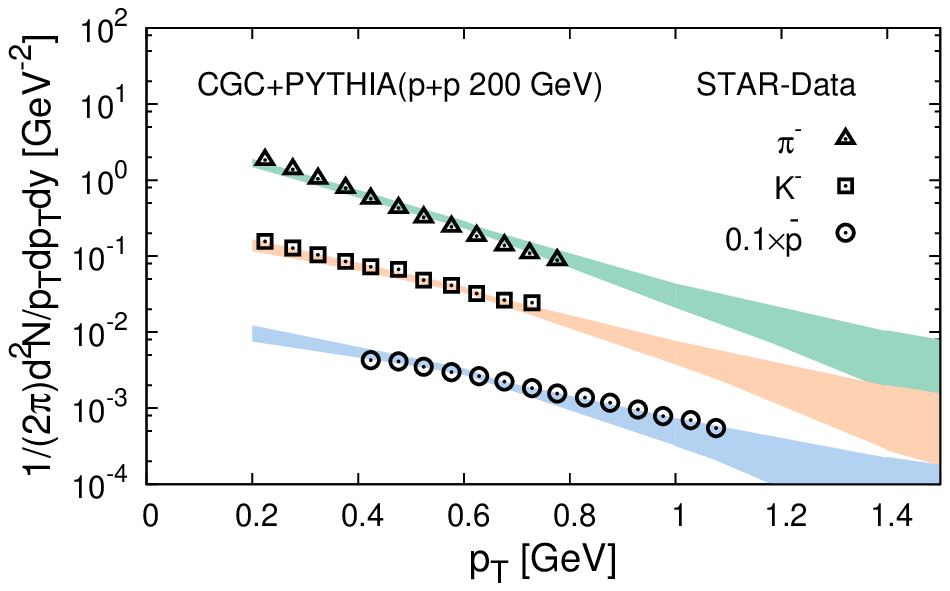}
\includegraphics[width=0.45\textwidth]{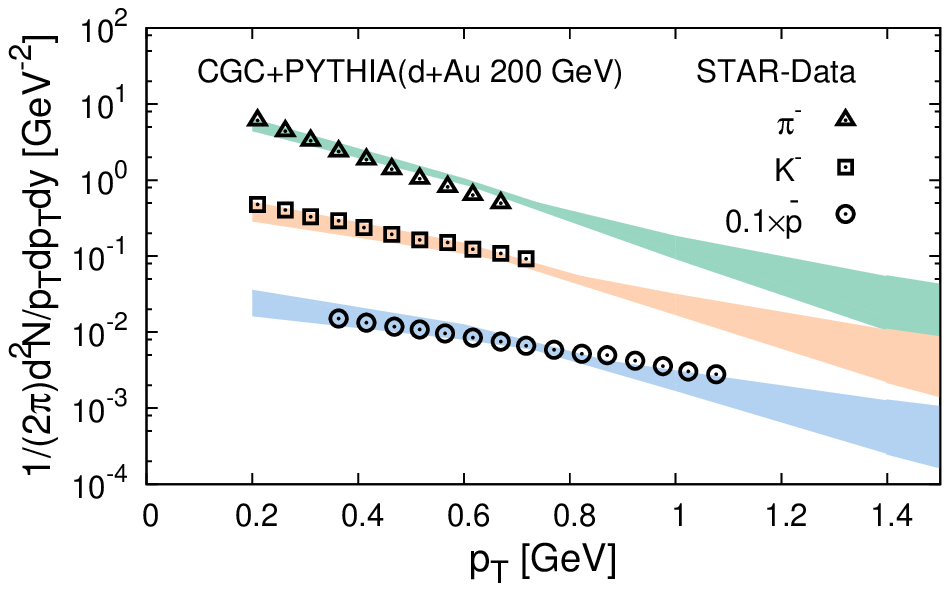}
\caption{(color online) Transverse momentum spectra of identified particles obtained from the CGC+PYTHIA framework (shown by bands) in 200 GeV p+p and d+Au collisions compared to the STAR data (shown by symbols).}
\label{fig_ptdist}
%\vspace{-10pt} 
\end{figure}

We now turn to more differential measurements. In Fig.\ref{fig_ptdist} we present the identified particle spectra for the two systems and compare them to the available data from STAR~\cite{Abelev:2008ez}. We present our results as shaded bands which already incorporate the systematic uncertainties in our calculations. These uncertainties include the variation of the evolution time for the Yang-Mills phase in the range of $\tau=0.4-0.6$ fm, the variation of the Lund string fragmentation parameters $a$ and $b$ within the range allowed by PYTHIA and the variation of $k_\perp^{\rm max}$ within a range of $8-15$ GeV. Within the uncertainties shown here, CGC+PYTHIA seems to provide a reasonable description of the identified particle spectra at RHIC. 
%
%The results shown here are obtained at fixed coupling. Additional uncertainties %arising in our calculations 
%
%from running coupling effects are estimated to be around $10-15\%$; they are dominant at high $p_{T}$ and not shown in Fig.\ref{fig_ptdist}. %Uncertainties in our framework grow at high $p_{T}$ where one expect to enter the regime of large-x. %Also, it must be noted that the kinematics of RHIC complicate data-model comparison at high $p_T$ due to large-x 

\begin{figure}
\includegraphics[width=0.45\textwidth]{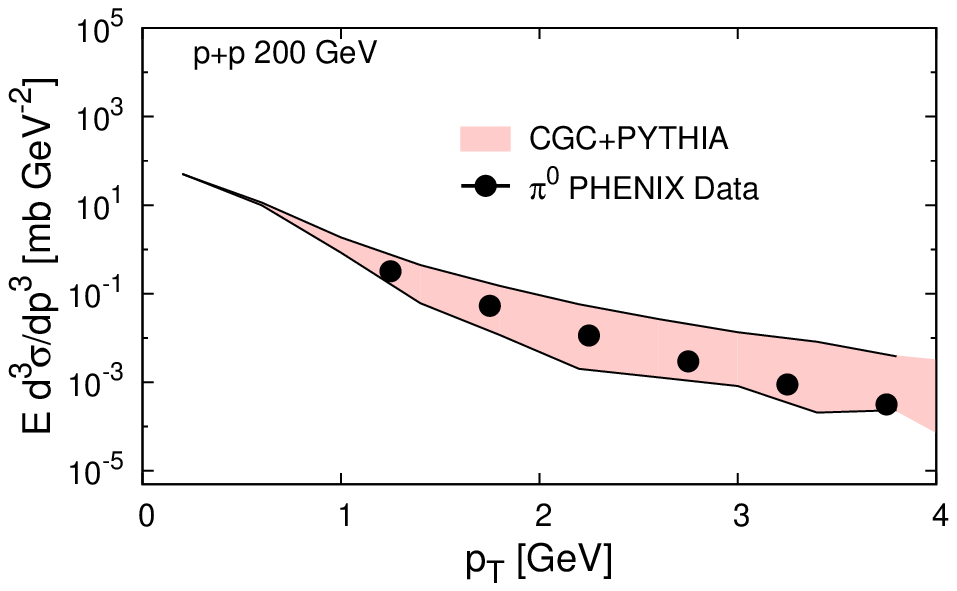}
\includegraphics[width=0.45\textwidth]{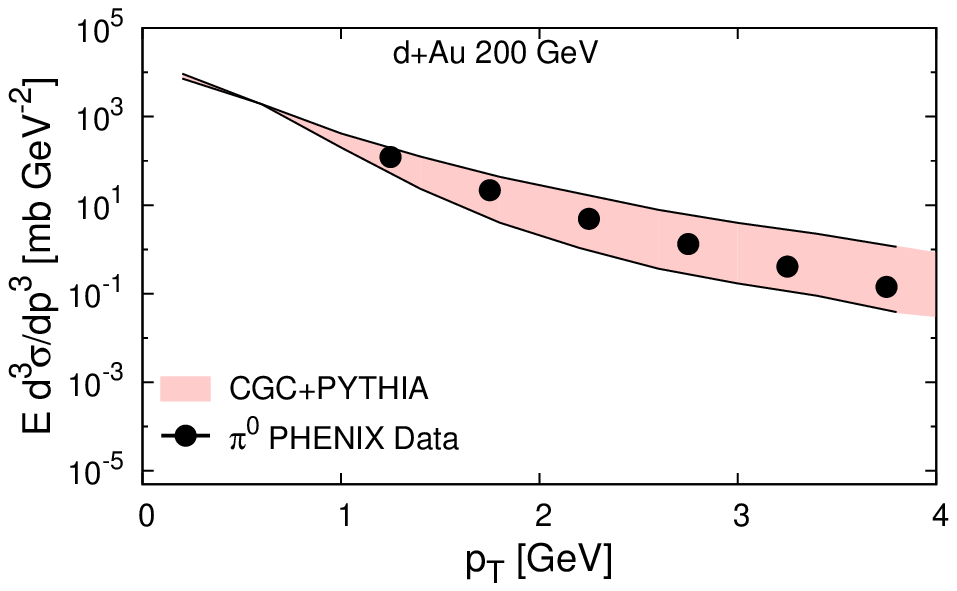}
\caption{(color online)  Transverse momentum dependence of the invariant cross section for $\pi^0$ in 200 GeV p+p and d+Au collisions compared to the PHENIX data from ~\cite{Adler:2006wg}.}
\label{fig_pi0dist}
%\vspace{-10pt} 
\end{figure}

In Fig.\ref{fig_pi0dist} we show the invariant cross section at mid-rapidity for $\pi^0$ within the pseudorapidity window of $|\eta|<0.3$ defined as $\sigma \times 1/(2\pi p_T)dN/dp_{T}dy$. We use $\sigma$ to be the inelastic cross section in p+p collisions $\sigma^{pp}_{\rm inel}$=42 mb and the total geometric cross section in d+Au collisions $\sigma^{\rm d+Au}_{\rm geo}$=2180 mb estimated by the MC-Glauber model calculations in~\cite{Alver:2008aq}. Once again we show CGC+PYTHIA calculations as bands that include different aforementioned sources of systematic uncertainties. 
Within the uncertainties the calculations agree well with the PHENIX data from~\cite{Adler:2006wg}.

\section{Summary}
%In this conference proceedings we apply the newly developed CGC+PYTHIA framework the study of bulk hadronic observables at RHIC
%In this contribution we present the very first study of small systems at RHIC using the newly developed CGC+PYTHIA framework. looks promising. 
In this contribution, we extend our previous work using the newly developed CGC+PYTHIA framework to p+p and d+Au collisions at RHIC. A first data-model comparison of the bulk hadronic observables such as the probability distribution of inclusive charged hadron multiplicity, the identified particle spectra, and invariant cross section of neutral pions looks promising. 
There are now several new results from the small system scan at RHIC. 
%There are new several results from the small collision systems scan at RHIC have accumulated; 
%
One particularly striking observation is the hierarchy of $v_{n}(p_T)$ in $p/d/He-Au$ collisions~\cite{Aidala:2018mcw}. Recent computations based on the CGC successfully describes such observations from initial state dynamics alone~\cite{Mace:2018vwq}. A similar hierarchy of $R_{p/d/He-Au}$ observed at RHIC~\cite{Sakaguchi:2016gec} also demands a first principles explanation. 
A future extension of our work will focus on more complex observables for the small collision systems scan at RHIC.
In addition to further exploration of small systems at RHIC and LHC, CGC+PYTHIA is also a promising framework for the phenomenology studies at a future EIC.
%~\cite{Aschenauer:2017jsk}. 

%We investigate several sources of systematic uncertainties, 
%
\section{Acknowlegement}
B. P. S., P. T., and R. V. are supported under DOE Contract No. DESC0012704. S. S. is supported by DOE Award No. DE-FG02-97ER41014. This research used resources of the National Energy Research Scientific Computing Center, which is supported by the Office of Science of the U.S. Department of Energy under Contract No. DE-AC02-05CH11231. %P.T. thanks T. Sj\"ostrand for useful discussions. 
We thank Kaushik Roy for a careful reading of the manuscript.

\bibliographystyle{elsarticle-num}
\bibliography{cymlundqm18.bib}






% \end{thebibliography}

\end{document}